# Solitons in a system of three linearly coupled fiber gratings

Arthur Gubeskys and Boris A. Malomed

Department of Interdisciplinary Studies
Faculty of Engineering, Tel Aviv University
Tel Aviv 69978, Israel

Abstract

We introduce a model of three parallel-coupled nonlinear waveguiding cores equipped with Bragg gratings (BGs), which form an equilateral triangle. The most promising way to create multi-core BG configuration is to use *inverted gratings*, written on internal surfaces of relatively broad holes embedded in a photonic-crystal-fiber matrix. The objective of the work is to investigate solitons and their stability in this system. New results are also obtained for the earlier investigated dual-core system. Families of symmetric and antisymmetric solutions are found analytically, extending beyond the spectral gap in both the dual- and tri-core systems. Moreover, these families persist in the case (strong coupling between the cores) when there is *no gap* in the system's linear spectrum. Three different types of asymmetric solitons are found (by means of the variational approach and numerical methods) in the tri-core system. They exist only inside the spectral gap, but asymmetric solitons with nonvanishing tails are found outside the gap as well. Stability of the solitons is explored by direct simulations, and, for symmetric solitons, in a more rigorous way too, by computation of eigenvalues for small perturbations. The symmetric solitons are stable up to points at which two types of asymmetric solitons bifurcate from them. Beyond the bifurcation, one type of the asymmetric solitons is stable, and the other is not. Then, they swap their stability. Asymmetric solitons of the third type are always unstable. When the symmetric solitons are unstable, their instability is oscillatory, and, in most cases, it transforms them into stable breathers. In both the dual- and tri-core systems, the stability region of the symmetric solitons extends far beyond the gap, persisting in the case when the system has no gap at all. The whole stability region of antisymmetric solitons (a new type of solutions in the tri-core system) is located outside the gap. Thus, solitons in multi-core BGs can be observed experimentally in a much broader frequency band than in the single-core one, and in a wider parameter range than it could be expected. Asymmetric delocalized solitons, found outside the spectral gap, can be stable too.

## 1. INTRODUCTION

Bragg gratings (BGs) are optical structures in which the refractive index is modulated with the spatial period equal half the wavelength of light propagating in it. The resonant modulation gives rise to mutual conversion of counter-propagating waves via the Bragg reflection. The BGs are widely used as optical filters, sensors and multiplexers [1]. Current applications exploit linear properties of the grating, namely, the presence of a gap in its spectrum, which is induced by the linear conversion of the counter-propagating waves. On the other hand, combination of this linear feature with intrinsic nonlinearity of materials in which BG can be written has attracted a great deal of attention in fundamental and applied research of the light transmission in BG-equipped structures. Nonlinearity gives rise to self-phase modulation (SPM) and cross-phase modulation (XPM) terms in the coupled-mode description of BGs. In particular, in the case when the BG is written on an optical fiber, the corresponding model is based on well-known equations [2]-[4],

$$iU_x + iU_t + V + \left(\frac{1}{2}|U|^2 + |V|^2\right)U = 0,$$

$$-iV_x + iV_t + U + \left(\frac{1}{2}|V|^2 + |U|^2\right)V = 0,$$

(1.1)

where $U$ and $V$ are amplitudes of the forward- and backward-propagating waves, $x$ is the coordinate along the fiber, the Bragg-reflection and XPM coefficients being normalized to be 1.

Analysis of stationary properties of this model led to prediction of optical bistability in nonlinear periodic structures in 1979 [2]. Later, a lot of work has been done on the theoretical [3]-[5] and experimental [6] study of gap solitons in this system. Gap solitons owe their existence to the balance between the SPM and XPM nonlinearity and BG-induced linear dispersion. The analytical form of the solitons is [3]-[4]

$$U(x,t;\psi,c) = A \operatorname{sech}(\xi - i\psi/2)\exp(i\theta),$$
$$V(x,t;\psi,c) = B \operatorname{sech}(\xi + i\psi/2)\exp(i\theta),$$

$$(A,B) = \pm\left(\frac{1\pm c}{1\mp c}\right)^{1/4}\sqrt{\frac{2(1-c^2)}{3-c^2}}\sin\psi,$$

$$\xi = \frac{x-ct}{\sqrt{1-c^2}}\sin\psi,$$

$$\theta = \frac{cx-t}{\sqrt{1-c^2}}\cos\psi - \frac{4c}{3-c^2}\tan^{-1}(|\cot(\psi/2)|\coth(\xi)),$$

(1.2)

Solutions belonging to this family depend on the parameters $\psi$ and $c$. The former one determines the amplitude, width, and central frequency of the soliton, and takes values from the interval $0 < \psi < \pi$, while the velocity $c$ is limited to $-1 < c < +1$.

Stability of the solutions (1.2) was first considered by means of a variational approximation in Ref. [7]. It was concluded that a part of the solitary-wave family (1.2) is unstable. The confirmation of the existence of an instability region in the $(\omega,c)$ plane of the solutions (1.2) was later obtained by means of rigorous numerical methods in the works [8]-[10] (see also Ref. [11]).

The study of gap solitons was then expanded to composite structures involving BGs. The simplest version of such a structure is a set of two linearly coupled parallel fibers carrying BGs. This system was investigated in Ref. [12]. It was shown that for some critical value of the linear-coupling coefficient, which depends on the energy of the soliton, obvious symmetric solitons lose their stability, bifurcating into stable

*asymmetric* ones. Stability analysis of both the symmetric and asymmetric soliton solutions was performed in Ref. [12] by means of direct simulations, and it was concluded that, when asymmetric solitons exist, they are always stable, while the symmetric ones are always unstable in the same case; on the other hand, when asymmetric solitons do not exist, the symmetric ones are always stable.

In this work, we investigate a triangular configuration of three linearly coupled identical BGs, which is the *most symmetric* possible composite structure consisting of waveguide gratings. In Ref. [13], solitons in a triangular configuration of ordinary nonlinear optical fibers were investigated, and solitons in a planar – rather than triangular – tri-core configuration were studied in detail in Ref. [14]. A system of three waves in the spatial domain, linearly coupled by a triple BG written on the surface of a planar nonlinear planar waveguide, was studied in Ref. [15]. This three-wave system, which is formally tantamount to a model with "1.5" cores in the temporal domain, gives rise to a rich family of solitons, which contains both regular gap solitons and stable "cuspons" and "peakons".

The consideration of the triangular three-core configuration of BGs is relevant for various reasons. First of all, this configuration makes it possible to perform direct switch of optical signals between any two. Generally, the study of nonlinear dynamical states in equilateral triangular configurations in various systems is a topic of fundamental interest for obvious symmetry reasons, see, e.g., Refs. [13] and [16]. In the context of the BG systems, tri-core configurations offer new possibilities to enhance functionality of the grating-based devices, which is an issue of considerable current interest, see a review [17]. We demonstrate that the transition from two to three cores drastically changes both the linear spectrum and soliton content of the system; for instance, instead of the single family of asymmetric solitons existing in the dual-core system, the tri-core one gives rise to three asymmetric families, two of which may be stable. Another essentially new feature of the tri-core system is the existence of a family of nontrivial antisymmetric solitons (in the dual-core model, symmetric and antisymmetric soliton families are tantamount to each other); the stability region of the antisymmetric solitons takes an unusual form, see below.

Besides that, we also report new results for dynamical states in dual-core nonlinear gratings. An essential result is that the existence and stability of symmetric solitons, in both the dual- and tri-core systems, continuously extend far *across the borders* of the spectral gap. This finding is of direct relevance to the experiment, as it suggests that the creation of stable solitons is possible in a frequency band and in a parameter region which are much broader than it could be expected *a priori*. Moreover, a broad stability region for the symmetric solitons is found in the case when the gap *does not exist* at all, hence the system cannot support ordinary gap solitons. In addition to that, it is found that the *whole* stability region of the antisymmetric solitons in the tri-core system is located outside the gap (and it also persists when the gap does not exist anymore).

The stability region of the symmetric solitons is limited to relatively small values of their amplitude. However, in most cases when they are unstable, the instability, which has an oscillatory character, does not destroy the solitons, but quickly transforms them into robust (numerically stable) *breathers*. The latter result is physically relevant too, as it suggests a possibility to experimentally look for such breathers in dual- and tri-core nonlinear gratings. Besides that, we demonstrate that stable *asymmetric* slightly delocalized solitons (the ones with nonvanishing tails) can also be found outside the gap, which makes the variety of stable dynamical states amenable to the experimental observation still broader. It is relevant to stress that all these stable states (solitons outside the gap, breathers, etc.) do not exist in the ordinary single-core gratings.

As concerns their actual realization, dual- and multi-core fiber gratings were a theoretical concept until the fabrication of a dual-core waveguide with the grating symmetrically written on both cores was reported in a very recent experimental work [18] (see Fig. 6 in that paper). In that work, the dual-core fiber grating was used as a basis for an add-drop telecommunications filter. The same technology should make the fabrication of the tri-core system quite feasible.

However, the most promising host medium in which dual- and tri-core gratings may be implemented is a photonic crystal fiber (PCF). Indeed, a guided mode can be easily localized inside the PCF in a layer of the width ~ 1 μm around a central hole of a relatively large diameter (~ 2 μm), see a review [19]. A natural way to apply the BG to this mode is to write the grating on the *internal surface* of the hole. Then, exactly the same model which describes a double- or tri-core fiber grating is also valid for the PCF matrix hosting a system of two or three broad holes which form an equilateral triangle, which is quite possible technologically. Note that fabrication and application of a PCF hosting two far separated waveguiding cores was reported very recently in Ref. [20] (see Fig. 1 in that paper); the same paper emphasizes that fabrication of various multi-core patterns in the PCF matrix is much easier than making similar structures composed of ordinary fibers. It is relevant to stress that that, although a combination of a PCF with BG was mentioned in some recent papers [21], the possibility to use *inverted* BGs, written on the internal surfaces of holes, and compose multi-core gratings of this type, were not yet considered, to our knowledge. Actually, applications of the inverted gratings in the PCF matrix may be much broader than just the creation of dual- and tri-core gratings.

In the context of making gap solitons in fiber gratings, the strength of the nonlinearity is a crucial issue, as the corresponding nonlinearity length must be no longer than ~ 1 cm [6]; in ordinary fibers, this makes it necessary to launch signals whose peak power is comparable to the optical-breakdown threshold. In the PCF setting, this problem is much easier to solve, as this medium may provide for an effective nonlinearity coefficient for the guided mode up to 100 $(W \cdot km)^{-1}$, i.e., 50 times as strong as in the ordinary fibers [22]. This is another advantage of using the PCF medium to create systems supporting gap solitons.

To conclude the introduction, it may be relevant to mention that another way to introduce a multi-component system in BGs is to use two polarizations of light (rather than multi-core structures), which was demonstrated experimentally in the temporal domain [23], and theoretically in the spatial one [24].

The paper is organized as follows. The formulation of the model is presented in section II. In section III, linear properties of the structure are considered. A gap in the system's spectrum is found, which outlines the region where gap solitons are expected to exist. Section IV presents families of exact symmetric and antisymmetric soliton solutions (as it was mentioned above, these soliton families extend across the gap's borders, and persist in the case when the system's spectrum has no gap at all). Section V investigates general asymmetric solitons. Three different species of them (on the contrary to the single family of asymmetric solitons in the dual-core system [12]) are predicted by means of the variational approximation, and are then found in a numerical form. Section VI deals with stability of the solitons. The stability of asymmetric solutions is tested by direct numerical simulations. For the above-mentioned exact symmetric solutions, *rigorous* stability analysis is performed within the framework of linearized equations for small perturbations. For the purpose of this consideration, the problem is formulated for $M$ symmetrically coupled Bragg gratings, so that it comprises the cases of both two ($M = 2$) and three ($M = 3$) linearly coupled cores. Conclusions are presented in section VII.

## 2. THE MODEL

We start by considering $M$ symmetrically coupled nonlinear gratings, where $M$ will take values 2 and 3. The electromagnetic field in the cores is assumed in the form

$$E_m(z,\tau) = E_{m+}(z,\tau)\exp(ik_0 z - i\omega_0 \tau) + E_{m-}(z,\tau)\exp(-ik_0 z - i\omega_0 \tau) + \text{c.c.}, \tag{2.1}$$

where the subscript $m = 1, ..., M$ is the number of the corresponding core, $z$ is the propagation distance, $\tau$ is time, $\omega_0$ and $k_0$ are the frequency and wave number of the carrier wave, and c.c. stands for the complex-conjugate expression.

Following the known technique of the coupled-mode theory, we derive evolution equations governing the slow evolution of the envelope fields in the cores:

$$i\frac{\partial E_{m+}}{\partial z} + i\frac{\partial E_{m+}}{\partial \tau} + \kappa E_{m-} + \gamma\left(|E_{m+}|^2 + 2|E_{m-}|^2\right)E_{m+} + C\sum_{k\neq m} E_{k+} = 0,$$

$$-i\frac{\partial E_{m-}}{\partial z} + i\frac{\partial E_{m-}}{\partial \tau} + \kappa E_{m+} + \gamma\left(|E_{m-}|^2 + 2|E_{m+}|^2\right)E_{m-} + C\sum_{k\neq m} E_{k-} = 0, \tag{2.2}$$

where $\kappa$ and $C$ are the Bragg reflectivity and the inter-core coupling constant, respectively, and the group velocity of light is set equal to 1 (the derivation is a straightforward combination of those for the single-core BG [5] and dual- or tri-core nonlinear couplers, see, e.g., Ref. [13]).

Applying the following normalizations to Eqs. (2.2),

$$E_{m+} = \sqrt{\frac{\kappa}{2\gamma}} U_m, \qquad z = \frac{x}{\kappa},$$

$$E_{m-} = \sqrt{\frac{\kappa}{2\gamma}} V_m, \qquad \tau = \frac{\overline{n}}{\kappa} t, \tag{2.3}$$

we cast them in a more convenient form:

$$iU_{mx} + iU_{mt} + V_m + \left(\frac{1}{2}|U_m|^2 + |V_m|^2\right)U_m + \lambda\sum_{k\neq m} U_k = 0, \quad (2.4)$$

$$-iV_{mx} + iV_{mt} + U_m + \left(\frac{1}{2}|V_m|^2 + |U_m|^2\right)V_m + \lambda\sum_{k\neq m} V_k = 0,$$

where the single remaining parameter is $\lambda = C/\kappa$. An essential difference of the tri-core model from its dual-core counterpart is the importance of the coupling-constant's sign: in the case of two coupled equations, one may always redefine $\lambda$ so that it is positive, while, in the case of three coupled equations, there is a real difference between positive and negative values of the coupling constant.

Throughout this paper, we consider only quiescent (zero-velocity) solitons, that can be looked for as

$$U_m = u_m(x)\exp(-i\omega t), \quad (2.5)$$
$$V_m = v_m(x)\exp(-i\omega t).$$

Substituting these expressions into Eqs. (2.4) leads to stationary equations,

$$\omega u_m + iu'_m + v_m + \left(\frac{1}{2}|u_m|^2 + |v_{m1}|^2\right)u_m + \lambda\sum_{k\neq m} u_k = 0, \quad (2.6)$$

$$\omega v_m - iv'_m + u_m + \left(\frac{1}{2}|v_m|^2 + |u_m|^2\right)v_m + \lambda\sum_{k\neq m} v_k = 0,$$

where the prime stands for $d/dx$. Forward- and backward propagating components of the quiescent solitons obey the symmetry constraints, $u = -v^*$. Then, Eq. (2.6) for the $m$-th core becomes

$$\omega u_m + iu'_m - u_m^* + \frac{3}{2}|u_m|^2 u_m + \lambda\sum_{k\neq m} u_k = 0. \quad (2.7)$$

Below, the stationary equations (2.7) and evolution equations (2.4) for two or three cores will be used to construct soliton solutions and investigate their stability. The main subject will be the new case of three cores, although the two-core model will be revisited two, in order to obtain some stability results in a more accurate form than it was done in Ref. [12].

Thus, for the tri-core configuration, the explicit forms of the propagation and stationary equations are, respectively,

$$iU_{1x} + iU_{1t} + V_1 + \left(\frac{1}{2}|U_1|^2 + |V_1|^2\right)U_1 + \lambda U_2 + \lambda U_3 = 0,$$

$$-iV_{1x} + iV_{1t} + U_1 + \left(\frac{1}{2}|V_1|^2 + |U_1|^2\right)V_1 + \lambda V_2 + \lambda V_3 = 0,$$

$$iU_{2x} + iU_{2t} + V_2 + \left(\frac{1}{2}|U_2|^2 + |V_2|^2\right)U_2 + \lambda U_1 + \lambda U_3 = 0,\qquad(2.8)$$

$$-iV_{2x} + iV_{2t} + U_2 + \left(\frac{1}{2}|V_2|^2 + |U_2|^2\right)V_2 + \lambda V_1 + \lambda V_3 = 0,$$

$$iU_{3x} + iU_{3t} + V_3 + \left(\frac{1}{2}|U_3|^2 + |V_3|^2\right)U_3 + \lambda U_2 + \lambda U_1 = 0,$$

$$-iV_{3x} + iV_{3t} + U_3 + \left(\frac{1}{2}|V_3|^2 + |U_3|^2\right)V_3 + \lambda V_2 + \lambda V_1 = 0,$$

and

$$\omega u_1 + iu_1' - u_1^* + \frac{3}{2}|u_1|^2 u_1 + \lambda u_2 + \lambda u_3 = 0,$$

$$\omega u_2 + iu_2' - u_2^* + \frac{3}{2}|u_2|^2 u_2 + \lambda u_1 + \lambda u_3 = 0,\qquad(2.9)$$

$$\omega u_3 + iu_3' - u_3^* + \frac{3}{2}|u_3|^2 u_3 + \lambda u_1 + \lambda u_2 = 0.$$

## 3. THE DISPERSION RELATION AND SPECTRAL GAP

Investigation of the linear spectrum can provide a clue to search for an existence range of gap solitons. Omitting nonlinear terms in Eqs. (2.8) and looking for a solution in the form

$$U_n = u_n \exp(ikx - i\omega t),$$
$$V_n = v_n \exp(ikx - i\omega t),\qquad(3.1)$$

we find the following branches of the dispersion relation:

$$\omega_{1,2} = -(M-1)\lambda \pm \sqrt{1+k^2},$$
$$\omega_{3,4} = \omega_{5,6} = \lambda \pm \sqrt{1+k^2},\qquad(3.2)$$

which are shown in Fig. 1. The dispersion dependences are written in the form which applies to both the dual-core ($M = 2$) and tri-core ($M = 3$) systems; in the former case, the branches $\omega_{5,6}$ are absent.

Gap solitons are expected to exist in the spectral gaps, where no linear propagation is possible. According to Eqs. (3.2), for the tri-core system the gap is

$$-(1-\lambda) < \omega < 1 - 2\lambda \qquad(3.3)$$

for positive $\lambda$, and

$$-(1+2\lambda) < \omega < 1+\lambda \tag{3.4}$$

for negative $\lambda$, while for the dual-core configuration the gap is $|\omega| < 1 - |\lambda|$. The gap is widest when $\lambda = 0$. With the increase of $|\lambda|$, the gap narrows, both intervals (3.3) and (3.4) shrinking to nil at $|\lambda| = 2/3$, similar to the dual-core's gap, which closes down at $|\lambda| = 1$.

In the case of two gratings, the gap is symmetric relative to the zero-detuning frequency, $\omega = 0$. On the other hand, in the tri-core system, this symmetry is absent, and the frequency $\omega = 0$ is completely pushed out from the spectral gap for $|\lambda| > 1/2$, while the gap still exists in the region $1/2 < |\lambda| < 2/3$. The gap region in the parametric space $(\omega, \lambda)$ of the tri-core system is shown in Fig. 2.

The expressions (3.3) and (3.4) define the genuine gap, i.e., an overlap between subgaps of all the dispersion branches. Regular gap solitons are expected to be found inside this gap. However, it is known that, in a system with at least two different branches of the dispersion relation, the so-called embedded solitons may exist inside one of the subgaps, being *embedded* in the continuous spectrum belonging to the other dispersion branch [25,26]. It will be shown below that the present system supports solitons both inside and outside the genuine gap. The embedded solitons are usually semi-stable (i.e., stable in the linear approximation, but nonlinearly unstable), and in some cases they turn out to be, in practical terms, completely stable objects [26].

## 4. EXACT SOLUTIONS

Some solitary-wave solutions of Eqs. (2.6) can be found in an exact analytical form. With respect to relations between the wave fields in different cores, these solutions may have a symmetric or antisymmetric form.

### A. Symmetric solutions

Symmetric solutions have identical fields $U$ and $V$ in all the cores. Looking for them in the form

$$U(x,t;\omega) = u(x;\omega)\exp(-i\omega t), \quad V(x,t;\omega) = v(x;\omega)\exp(-i\omega t), \tag{4.1}$$

it is straightforward to see that they can be expressed as

$$u(x;\omega,\lambda) = u_B(x;\omega + (M-1)\lambda), \quad v(x;\omega,\lambda) = v_B(x;\omega + (M-1)\lambda), \tag{4.2}$$

(recall $M$ is the number of cores, which takes values 2 or 3), where $u_B(x;w)$ and $v_B(x;w)$ are the standard stationary solutions (1.2) for the gap solitons with $c = 0$ and the frequency $\cos\psi = \omega + (M-1)\lambda$ in the single-core nonlinear BG. These solutions exist in the frequency interval $-1 < \cos\psi < +1$, which is the corresponding spectral gap, hence the exact symmetric solutions can be found in the interval

$$-1 < \omega + (M-1)\lambda < 1, \tag{4.3}$$

i.e., precisely in the subgap of the dispersion branches $\omega_{1,2}(k)$ defined by Eqs. (3.2). In the case of the tri-core system, $M = 3$, the region (4.3) is the interior of the negative-slope stripe in Fig. 2. Thus, the exact symmetric-soliton solutions exist not only inside the gap proper, but also (as embedded solitons) in the whole subgap (4.3).

It is also important to notice that the family of the symmetric solitons is present even in the cases $|\lambda|>1$ and $|\lambda|>2/3$, for $M = 2$ and $M = 3$, respectively, when the genuine gap *does not exist* at all, hence the system supports no regular gap solitons, while the continuous family of the embedded symmetric solitons is available.

### B. Antisymmetric solutions

The existence of the antisymmetric solution in the dual-core model is obvious [12]. In fact, they are tantamount to the symmetric solitons with $\lambda$ replaced by $-\lambda$. In the tri-core system, this symmetry is absent, and an antisymmetric soliton is an independent solution. It has components with opposite signs in two cores, and zero in the third one: for instance, $(U,V)_1 = 0$, and $(U,V)_2 = -(U,V)_3 \equiv (U,V)$. The existence of this solution is possible as the linear terms generated by the first and second fields in the equation for the third core exactly cancel each other.

In the non-empty cores, the stationary field for the antisymmetric solution takes the form

$$u(x;\omega,\lambda) = u_B(x;\omega-\lambda), \quad v(x;\omega,\lambda) = v_B(x;\omega-\lambda), \qquad (4.4)$$

where $u_B(x;\omega)$ and $v_B(x;\omega)$ are the standard single-core solutions (1.2) with $c = 0$ and frequency $\cos\psi = \omega - \lambda$, cf. Eqs. (4.2). Accordingly, the antisymmetric solutions exist in the interval

$$-1 < \omega - \lambda < 1, \qquad (4.5)$$

cf. Eq. (4.3). This region is exactly the subgap of the dispersion branches $\omega_{3,4}$ and $\omega_{5,6}$ from Eqs. (3.2), and it fills the stripe with the positive slope in Fig. 2. Thus, as well as the symmetric solitons, the antisymmetric ones exist not only inside the gap proper, but also in the whole subgap (the one transversal to that which supports the symmetric solitons). Beyond the gap's borders, the antisymmetric solitons are embedded ones. Lastly, we notice that, as well as their symmetric counterparts, the antisymmetric solitons keep to exist (as embedded solitons) even in the case when the true gap is absent in the system.

### 5. ASYMMETRIC SOLITONS: THE VARIATIONAL APPROXIMATION AND NUMERICAL RESULTS

In order to classify more general soliton solutions, it is natural to start the analysis with the case of a small inter-core coupling constant $\lambda$ (similar to how it was done for the dual-core model in Ref. [12]). In the limit $\lambda \to 0$, equations for different cores decouple, and the following solutions can be identified [recall $(u,v)_B$ stands for the usual single-core soliton with the zero velocity]:

- Symmetric, with $(u,v)_1 = (u,v)_2 = (u,v)_3 = (u,v)_B$ ;
- Antisymmetric, $(u,v)_1 = -(u,v)_2 = (u,v)_B$, $(u,v)_3 = 0$ ;
- Asymmetric solution, *type I*: $(u,v)_1 = (u,v)_B$, $(u,v)_2 = (u,v)_3 = 0$ ;
- Asymmetric solution *type II*: $(u,v)_1 = (u,v)_B$, $(u,v)_2 = (u,v)_3 = -(u,v)_B$;

- Asymmetric solution, *type III*: $(u,v)_2 = (u,v)_3 = (u,v)_B$, $(u,v)_1 = 0$.

It is relevant to mention that a similar approach, in which different types of solutions are identified on the basis of obvious forms available in the decoupled limit, is very efficient in classification of various soliton-like states in models of dynamical lattices, see, e.g., Ref. [27] and references therein. Another essential remark is that, in the dual-core system, the same limiting case, $\lambda \to 0$, gives rise to a single type of asymmetric solutions, with $(u,v)_1 = (u,v)_B$, $(u,v)_2 = 0$.

For the first two species (symmetric and antisymmetric solitons), exact solutions valid for finite $\lambda$ were given in the previous section. To find the stationary solutions of all the types, we first applied the variational approximation (VA) to Eqs. (2.9); then, the solutions were sought for in a direct numerical form.

First, we present analytical results provided by VA. The tri-core stationary equations (2.9) can be derived from the Lagrangian density, (5.1)

$$L = Q_B(u_1) + Q_B(u_2) + Q_B(u_3) + 2\lambda \operatorname{Re}\{u_1 u_2^* + u_1 u_3^* + u_2 u_3^*\}$$

where

$$Q_B(u) \equiv \omega |u|^2 + \frac{3}{4}|u|^4 - \frac{1}{2}u^2 - \frac{1}{2}(u^*)^2 - \operatorname{Im}(u'u^*)$$ (5.2)

is the Lagrangian density for the single-core model. Next, following Ref. [12], where VA was applied to soliton solutions of the dual-core model and yielded quite accurate results, we adopt the following *ansatz*,

$$u_n = A_n \operatorname{sech}(\mu x) + iB_n \sinh(\mu x)\operatorname{sech}^2(\mu x),$$ (5.3)

where the same width $\mu^{-1}$ is assumed for all the three components, while the amplitudes $A_n$ and $B_n$ may be different for different $n$. The *effective Lagrangian*, which is obtained by the substitution of the ansatz (5.4) in the density (5.1) and integration, is displayed in Appendix. Varying the effective Lagrangian (A.1) with respect to $A_n$, $B_n$, and $\mu$ yields a system of algebraic equations (A.2), which are also written in Appendix.

Equations (A.2) were solved by means of the Newton-Raphson method. Figures 3, 4 and 5 show amplitudes $A_n$ of two different components for solitons of all the types defined above, as found from the VA at $\omega = 0$, $\omega = 0.5$, and $\omega = 0.8$, respectively. Only two amplitudes are shown because the third one is always equal to one of them, and the symmetric solution is represented by a single curve, as it has a single amplitude.

As is seen from Figs. 3 – 5, branches of asymmetric solutions of the types I and III are generated from the symmetric-soliton branch by pitchfork bifurcations. Further, the following general conclusions can be drawn from the diagrams:

- Symmetric solutions bifurcate twice: at a larger value of $\lambda$ into Type-I asymmetric solutions, and at a slightly smaller $\lambda$ into Type-III solitons. The two bifurcation points tend to come closer as $\omega$ become larger.
- The region of existence of the asymmetric solitons of the types I and III becomes smaller as $\omega$ increases.
- The type-III solution exists only inside the genuine spectral gap (recall it was defined above as the overlap between the two subgaps).
- The type-I solution also exists inside the genuine gap only, unless $\omega = 0$ (probably, at small finite values of $\omega$ this solution may exist outside the genuine gap).
- The type-II solution exists both inside and outside the genuine gap.

Next, the relaxation method, based on the so-called sinc-collocation technique, was employed to obtain numerically exact stationary soliton solutions, using the prediction produced by the VA as an initial guess. Inside the genuine gap, the variational results match their numerical counterparts quite well, which is illustrated by Fig. 6.

On the other hand, direct attempts to generate asymmetric solitons outside of the genuine gap in the numerical form, starting with the VA-predicted initial guess (in those cases when the VA does predict asymmetric solitons outside the gap), have failed. Instead, the numerical algorithm produces delocalized solitons (alias "quasi-solitons") with small nonvanishing oscillatory tails, which is a direct consequence of the fact that the solution does not belong to the genuine gap. A definite conclusion verified by the numerical method is that regular asymmetric solitons exist *only inside* the spectral gap, while all the asymmetric solitons found outside the gap are delocalized ones. In the latter case, the central core of the delocalized solitons is found to be quite close to the corresponding VA-predicted shape, in accordance with the principle that VA, even if it misses the existence of the tail, is able to correctly describe the soliton's core [28].

Typical examples of the delocalized soliton, found outside the gap, are displayed in Fig. 7. An unexpected feature found from the numerical results is that the amplitude of the nonvanishing tail *increases* as one approaches the gap's border along the branches of the type-I or type-II delocalized solitons. Hitting the gap edge, the delocalized solitons disappear, while a regular (truly localized) asymmetric solitons of the types I and II show up. In fact, the central body of the quasi-solitons is described by the VA quite accurately.

## 6. STABILITY

After having found basic types of stationary soliton solutions in the model, it is necessary to analyze their stability. Direct simulations of the evolution equations is the most commonly used method for determining the stability of solitons, which actually corresponds to the way the physical experiment is run. However, more rigorous information on the stability is furnished by numerical computation of the corresponding eigenvalues within the framework of linearized equations for small perturbations (for the single-core Bragg-grating model, this was done in Refs. [8-10]). In particular, it may happen that direct simulations may sometimes manifest a weak instability which is an artifact of the numerical scheme, while the soliton is stable (in some cases considered in Ref. [12], the weak instability of symmetric solitons in the dual-core BG model was actually the artifact).

Below, we display results for the stability of solitons in the present models (both dual-core and tri-core ones), obtained by means of both direct simulations and eigenvalue computation. Direct simulations were used to investigate the stability of asymmetric solitons (for which numerical computation of the eigenvalues is a technically difficult problem), and to delineate a stability region of symmetric solitons. Then, we applied the more rigorous eigenvalue-based procedure to the symmetric solutions.

### A.      Direct simulations

Systematic direct simulations of the soliton stability have resulted in the following conclusions for the symmetric, antisymmetric, and asymmetric solitons in the tri-core system:

- Symmetric solitons are stable before the bifurcations occur. After passing (in the direction of decreasing $\lambda$, see Figs. 3-5) the bifurcation points at which the type-I and type-III asymmetric solitons are born, the symmetric-soliton branch becomes unstable. Since two bifurcation points may be quite close, it is sometimes difficult to determine exactly where the symmetric soliton loses its stability. On the other hand, *stable* (at least, in the linear approximation) symmetric solitons are also found *outside the gap*, see Fig. 10 below. The implication of the latter result for the experiment is the great enlargement of the parameter region (in particular, of the corresponding frequency band) in which stable solitons can be sought for. Moreover, stable symmetric solitons are found in the case $|\lambda| > 2/3$, when the gap *does not exist* at all in the tri-core system.
- The antisymmetric solution with one empty core is always unstable inside the spectral gap, and it is unstable too when it exists outside the gap for positive $\lambda$, see Figs. 4 and 5. However, at relatively large negative values of the coupling constant $\lambda$, the soliton of this type is *stable* (this situation occurs, e.g., in the case shown in Fig. 3; recall that, unlike the dual-core system, in the tri-core one the sign of $\lambda$ is a nontrivial ingredient). Nontrivial purport of this finding is that stable antisymmetric solitons (as well as symmetric ones, see above) are possible *outside the gap*, including the case when the gap *does not exist*. An example of stable evolution of such an asymmetric soliton is displayed in Fig. 8 for $\omega = 0$ and $\lambda = -0.8$ (notice that this stable antisymmetric soliton is found in the case when the true gap does not exist at all, as $|\lambda| > 2/3$ in this case).
- The type-III asymmetric soliton is stable as it is generated by the bifurcation from the symmetric soliton at positive $\lambda$. On the other hand, the type-I asymmetric solitons appear as unstable ones after the corresponding bifurcation. Following the branches of the type-I and type-III solitons towards negative values of $\lambda$, *stability exchange* between them was found. For example, the type-III soliton is stable at $\omega = 0$ and $\lambda = 0.2$, and its type-I counterpart is unstable in the same case, while for $\omega = 0$ and $\lambda = -0.2$ the character of the stability is exactly opposite. Figure 9 displays typical examples of stable type-I and type-III asymmetric solitons.
- Type-II asymmetric solitons are *always unstable*. Quasi-solitons (delocalized ones) of the same type are always unstable too.

- Type-I delocalized solitons (quasi-solitons) existing outside the gap are *stable* for relatively large negative values of $\lambda$. In particular, the stability was observed, at $\omega = 0$, for $\lambda$ taking values between $-0.6$ and $-0.8$. As well the stability of the truly localized antisymmetric solitons found outside the gap, the stability of the slightly delocalized asymmetric ones is a prediction of significance to the experiment, as it opens a way to look for stable patterns in a much larger parameter region than it could be expected.

**B.    The eigenvalue analysis**

In this paper, we present results of computation of the stability eigenvalues only for the symmetric solitons, as in other cases the problem is really difficult, and it should be a subject for a separate work. It is relevant first to recapitulate the way the eigenvalue analysis was performed for the single-core model in Ref. [8]. To derive linearized equations for small perturbations around the stationary symmetric solitons, the perturbed solutions were taken in the form

$$
\begin{aligned}
U &= (u_B(x;\omega) + \varepsilon_1(x)\exp(i\alpha t))\exp(-i\omega t), \\
V &= (v_B(x;\omega) + \varepsilon_2(x)\exp(i\alpha t))\exp(-i\omega t), \\
U^* &= (u_B^*(x;\omega) + \varepsilon_3(x)\exp(i\alpha t))\exp(i\omega t), \\
V^* &= (v_B^*(x;\omega) + \varepsilon_4(x)\exp(i\alpha t))\exp(i\omega t),
\end{aligned}
\quad (6.1)
$$

where $\alpha$ is the eigenfrequency (stability eigenvalue), and $(u,v)_B(x;\omega)$ are taken as per the exact solution (1.2). Then, eigenmodes of the small perturbations are solutions to the following linear system,

$$
\left[ i\begin{pmatrix} \sigma_3 & 0 \\ 0 & -\sigma_3 \end{pmatrix}\frac{\partial}{\partial x} + \begin{pmatrix} \sigma_1 & 0 \\ 0 & \sigma_1 \end{pmatrix} + \omega\begin{pmatrix} \sigma_0 & 0 \\ 0 & \sigma_0 \end{pmatrix} + Q(u_B, v_B) \right]\varepsilon = \alpha\begin{pmatrix} \sigma_0 & 0 \\ 0 & -\sigma_0 \end{pmatrix}\varepsilon,
\quad (6.2)
$$

where $\varepsilon$ is the column composed of $\varepsilon_1,...,\varepsilon_4$, $\sigma_n$ are the Pauli matrices, and the matrix $Q$ is

$$
Q(u_B, v_B) \equiv \begin{pmatrix}
|u_B|^2 + |v_B|^2 & v_B^* u_B & \tfrac{1}{2}u_B^2 & u_B v_B \\
u_B^* v_B & |u_B|^2 + |v_B|^2 & u_B v_B & \tfrac{1}{2}v_B^2 \\
\tfrac{1}{2}u_B^{*2} & u_B^* v_B^* & |u_B|^2 + |v_B|^2 & u_B^* v_B \\
u_B^* v_B^* & \tfrac{1}{2}v_B^{*2} & v_B^* u_B & |u_B|^2 + |v_B|^2
\end{pmatrix}
\quad (6.3)
$$

Equation (6.2) can then be rewritten as

$$
A(u_B, v_B, \omega)\varepsilon = \alpha J\varepsilon, \quad (6.4)
$$

defining $A$ and $J$ appropriately.

Equation (6.4) sets the eigenvalue problem. In Ref. [8], it was shown that, in the single-core model, there is a critical value $\omega_{cr}$ of the frequency of the unperturbed soliton, such that for $\omega > \omega_{cr}$ all the eigenvalues $\alpha$ are real [i.e., there is no instability, see Eqs. (6.1)]. In fact, $\omega_{cr}$ found in Ref. [8] is very close to zero (virtually the same critical value was found earlier in Ref. [7] by means of the VA).

In the case of the symmetric solitons in the dual- or tri-core system, we assume perturbations of the form

$$\begin{aligned}
U_n &= \left(\hat{u}_n(x;\omega,\lambda) + \varepsilon_1^n(x)\exp(i\alpha t)\right)\exp(-i\omega t), \\
V_n &= \left(\hat{v}_n(x;\omega,\lambda) + \varepsilon_2^n(x)\exp(i\alpha t)\right)\exp(-i\omega t), \\
U_n^* &= \left(\hat{u}_n^*(x;\omega,\lambda) + \varepsilon_3^n(x)\exp(i\alpha t)\right)\exp(i\omega t), \\
V_n^* &= \left(\hat{v}_n^*(x;\omega,\lambda) + \varepsilon_4^n(x)\exp(i\alpha t)\right)\exp(i\omega t),
\end{aligned} \qquad (6.5)$$

where $\hat{u}_n(x;\omega,\lambda)$ and $\hat{v}_n(x;\omega,\lambda)$ are the unperturbed symmetric solitary-wave solutions, like the one given by Eq. (4.1). Introducing the perturbation of the form (6.5) into Eq. (2.4), defining $\boldsymbol{\varepsilon}^n = [\varepsilon_1^n, \varepsilon_2^n, \varepsilon_3^n, \varepsilon_4^n]^T$, and making use of the definitions (6.3) and (6.4), we arrive at the new eigenvalue problem,

$$\begin{bmatrix} A(\hat{u}_1,\hat{v}_1,\omega) & \cdots & \lambda\sigma_0 \\ \cdots & \cdots & \cdots \\ \lambda\sigma_0 & \cdots & A(\hat{u}_M,\hat{v}_M,\omega) \end{bmatrix} \begin{bmatrix} \varepsilon^1 \\ \vdots \\ \varepsilon^M \end{bmatrix} = \alpha \begin{bmatrix} J & \cdots & 0 \\ \cdots & \cdots & \cdots \\ 0 & \cdots & J \end{bmatrix} \begin{bmatrix} \varepsilon^1 \\ \vdots \\ \varepsilon^M \end{bmatrix}, \qquad (6.6)$$

where $\sigma_0$ is the unity matrix.

Since the system is conservative, the stability may only be neutral, with all the eigenvalues being real. The actual objective is to solve the eigenvalue problem (6.6) for $M = 2$ and 3, varying the values of $\lambda$ and $\omega$, in order to identify a stability region in the plane $(\lambda,\omega)$ where all the eigenvalues are purely real.

For the symmetric solitons, Eqs. (6.6) can be radically simplified (while for asymmetric solitons they are very involved, that is why the eigenvalue analysis is not developed here for them). Indeed, by means of a linear transformation, Eqs. (6.6) for the symmetric soliton are cast in the form

$$\begin{bmatrix} A(\hat{u},\hat{v},\omega)+(M-1)\lambda\sigma_0 & 0 & \cdots & 0 \\ 0 & A(\hat{u},\hat{v},\omega)-\lambda\sigma_0 & \cdots & 0 \\ \cdots & \cdots & \cdots & \cdots \\ 0 & 0 & \cdots & A(\hat{u},\hat{v},\omega)-\lambda\sigma_0 \end{bmatrix} \begin{bmatrix} \sum_m \varepsilon^m \\ \varepsilon^1 - \varepsilon^2 \\ \vdots \\ \varepsilon^1 - \varepsilon^M \end{bmatrix}$$

$$= \alpha \begin{bmatrix} J & 0 & \cdots & 0 \\ 0 & J & \cdots & 0 \\ \cdots & \cdots & \cdots & \cdots \\ 0 & 0 & \cdots & J \end{bmatrix} \begin{bmatrix} \sum_m \varepsilon^m \\ \varepsilon^1 - \varepsilon^2 \\ \vdots \\ \varepsilon^1 - \varepsilon^M \end{bmatrix}, \qquad (6.7)$$

in which the system is effectively decoupled, consequently the full set of the eigenvalues is the union of sets obtained by solving each partial equation in (6.7) separately. The decoupled equations can be further simplified, making use of the property following from the definition (6.4),

$$A(\hat{u}(x;\omega,\lambda),\hat{v}(x;\omega,\lambda),\omega)+n\lambda\sigma_0 = A(\hat{u}(x;\omega,\lambda),\hat{v}(x;\omega,\lambda),\omega+n\lambda), \tag{6.8}$$

where $n$ is an arbitrary integer. Thus we obtain

$$\begin{bmatrix} A(\hat{u}_B(x;\omega+(M-1)\lambda),\hat{v}_B(x;\omega+(M-1)\lambda),\omega+(M-1)\lambda) & 0 \\ 0 & A(\hat{u}_B(x;\omega+(M-1)\lambda),\hat{v}_B(x;\omega+(M-1)\lambda),\omega-\lambda) \end{bmatrix} \tilde{\boldsymbol{\varepsilon}} \tag{6.9}$$

$$= \alpha \begin{bmatrix} J & 0 \\ 0 & J \end{bmatrix} \tilde{\boldsymbol{\varepsilon}},$$

where $\tilde{\boldsymbol{\varepsilon}} \equiv \left[ \sum_m \varepsilon^m , \varepsilon^1 - \varepsilon^2 \right]^T$.

Actually, the first decoupled equation implied in (6.9) has already been solved in the context of the single-core stability problem [6,8], which yields stable eigenvalues for $\omega+(M-1)\lambda > \omega_{cr}$, and unstable ones for $\omega+(M-1)\lambda < \omega_{cr}$.

Thus, the symmetric solitons are definitely unstable in the region $\omega < \omega_{cr} - (M-1)\lambda$. The actual character of the region $\omega > \omega_{cr} - (M-1)\lambda$, where the above result does not produce instability, is determined by eigenvalues of the second decoupled equation in (6.9). With the definitions

$$\omega' = \omega + (M-1)\lambda, \beta \equiv -M\lambda, \tag{6.10}$$

we arrive at a modified eigenvalue problem

$$A(u_B(x;\omega'),v_B(x;\omega'),\omega'+\beta)\boldsymbol{\varepsilon} = \alpha J\boldsymbol{\varepsilon}. \tag{6.11}$$

Eigenvalues generated by Eq. (6.11) were found numerically, using the above-mentioned sinc-collocation technique, with the sinc basis augmented by the sine and cosine functions, to account for nonvanishing periodic tails of the eigenfunctions at $x \to \pm\infty$ (an extensive description of the sinc techniques can be found in Ref. [29]). The computation of the eigenvalues was performed on a dense grid in the space $(\omega,\beta)$. Points at which the symmetric solitons were thus found to be completely stable are marked in Fig. 10 for both the dual-core and tri-core systems, $M = 2$ and 3 (as it was mentioned above, in the dual-core system antisymmetric solitons are tantamount to their symmetric counterparts, with $\lambda$ replaced by $-\lambda$, therefore in Fig. 10 the stability regions for the symmetric and antisymmetric solitons in the dual-core system are mirror images of each other).

The diagonal lines with the negative slope show the boundary of the existence of the symmetric solitons. A conclusion clearly suggested by Fig. 10 is that the symmetric solitons are stable close to the upper boundary of their existence region, which corresponds to small-amplitude symmetric solitons. Another important inference is that (as it was already mentioned above) the stability region of the symmetric solitons extends far beyond the borders of the spectral gap, and, moreover, stable symmetric solitons are found in the case when the gap does not exist ($|\lambda| > 1$ for $M = 2$, and $|\lambda| > 2/3$ for $M = 3$). Thus, the symmetric solitons form a *continuous family* of *embedded solitons* [25,26], which are stable in the linear approximation, even in the case when regular gap solitons do not exist at all.

Another relevant issue is to understand what will happen with the symmetric soliton when it is subject to the instability. Our computations clearly demonstrate that, in both the dual- and tri-core systems, the destabilization of the symmetric solitons occurs, with the increase of the soliton's amplitude, through the emergence of a pair of complex-conjugate eigenvalues, hence the instability is oscillatory. In accordance with that, direct simulations show that, most typically, the growth of unstable perturbations quickly saturates, and, as a result, the unstable static soliton turns into a *breather*, which features persistent intrinsic vibrations. To illustrate this generic scenario, in Fig. 11 we display a typical example of the destabilization of the symmetric soliton in the dual-core system. The destabilization sets in at $\omega = 0.86$ for $\beta = -0.4$ [i.e., $\lambda = 0.2$; the parameters are defined as per Eqs. (6.10)]. Direct simulations of the same system, displayed in Fig. 12, confirm that the development of the instability transforms the unstable symmetric soliton into a breather. Nevertheless, in rarer cases (at some other values of the parameters), the instability could completely destroy the symmetric soliton.

This generic result (the transformation of unstable symmetric solitons into stable breathers) is of obvious relevance to the experiment, as it suggests a possibility to observe the breathers in the dual- and tri-core nonlinear gratings. The only necessary condition for the creation of a breather is to start with an unstable soliton having a sufficiently large amplitude, i.e., launching a pulse carrying sufficiently large energy, which is not difficult [6].

Lastly, we dwell on antisymmetric solitons in the tri-core system. In this case, the eigenvalue problem (6.6) can be simplified too. Eventually, the full stability problem splits into two decoupled ones. This time, they take the form of

$$\begin{bmatrix} A(0,0,\omega) & \lambda\sigma_0 \\ \lambda\sigma_0 & A(\hat{u}_B(x;\omega-\lambda),\hat{v}_B(x;\omega-\lambda),\omega) \end{bmatrix} \begin{bmatrix} \varepsilon^1 \\ \varepsilon^2 \end{bmatrix} = \alpha \begin{bmatrix} J & 0 \\ 0 & J \end{bmatrix} \begin{bmatrix} \varepsilon^1 \\ \varepsilon^2 \end{bmatrix}, \tag{6.11}$$

$$A(\hat{u}_B(x;\omega-\lambda),\hat{v}_B(x;\omega-\lambda),\omega-\lambda)\varepsilon = \alpha J\varepsilon. \tag{6.12}$$

It follows from the separating equation (6.12) that the antisymmetric soliton is definitely unstable in the region $\omega < \omega_{cr} + \lambda$. The stability in the remaining region, $\omega > \omega_{cr} + \lambda$, is determined by two coupled equations (6.11). We do not present results of the full investigation of these equations here, which is a technically hard problem. Recall that direct simulations reported in the previous subsection have demonstrated that the antisymmetric solitons in the tri-core system may be stable outside the gap (or in the case when the gap does not exist at all) at sufficiently large negative $\lambda$.

### 7. CONCLUSION

In this work, we have introduced a model of three linearly coupled nonlinear cores equipped with Bragg gratings (BGs), which form a triangle. The most promising way to create tri-core BG configurations (as well as dual-core ones) is to use *inverted gratings* written on internal surfaces of relatively broad holes in the photonic-crystal-fiber setting. The investigation of solitons in this model and their stability is an issue of interest in its own right, and offers applications to the design of highly functional photonic devices, as well as to the study of photonic crystals combined with BGs. We have also revisited the earlier studied dual-core system, obtaining new results for it.

Families of symmetric and antisymmetric solutions were found analytically. They continuously extend across borders of the spectral gap, in both the dual- and tri-core systems; moreover, these families persist in the case when the system's linear spectrum has no gap at all, hence the system cannot support ordinary gap solitons. Apart from that, three different types of asymmetric solitons were found by means of the variational approach and numerical methods. Asymmetric solitons of all the types exist only inside the spectral gap, but slightly delocalized asymmetric solitons with nonvanishing tails are found outside the gap.

Stability of all the solitons was explored in direct simulation of the evolution equations. The stability of symmetric solitons was also studied in a more rigorous way, by computation of stability eigenvalues for small perturbations. The results show that the symmetric solitons are stable up to bifurcation points, at which they give birth to two types of asymmetric solitons. Beyond the bifurcation, one type of the asymmetric solitons is initially stable, while the other is not. Later, they swap their (in)stability. The third type of the asymmetric solitons is always unstable. If the symmetric solitons are unstable, their instability has oscillatory character. In most cases, it does not destroy them, but rather transforms into robust breathers.

It has been found that the stability region of the symmetric solitons, in both the dual- and tri-core systems, extends far beyond the spectral gap; moreover, a broad stability region of the symmetric solitons is found in the case (strong coupling between the cores) when there is no gap in the linear spectrum of the systems. In addition, the whole stability region of antisymmetric solitons, which constitute a new type of solutions in the tri-core system, is located outside the gap (and it also persists when the gap is absent). These findings suggest that solitons in the multi-core BGs can be observed experimentally in a much broader frequency band than in their single-core counterpart, and in a much wider parameter region than it could be expected *a priori*. Stable asymmetric delocalized solitons were also found outside the spectral gap, additionally extending the variety of nonlinear states amenable to the experimental observation in the multi-core BGs.

## ACKNOWLEDGEMENT

We appreciate useful discussions of issues concerning the soliton stability with Andreas Mayer.

## APPENDIX

The effective Lagrangian produced by the ansatz (5.3) is

$$\int_{-\infty}^{\infty} L(x)dx = \frac{1}{\mu}\left(A_1^4 + A_2^4 + A_3^4\right) + \frac{3}{35\mu}\left(B_1^4 + B_2^4 + B_3^4\right) + \frac{2}{5\mu}\left(A_1^2 B_1^2 + A_2^2 B_2^2 + A_3^2 B_3^2\right)$$
$$-\frac{2}{\mu}(1-\omega)\left(A_1^2 + A_2^2 + A_3^2\right) + \frac{2}{3\mu}(1+\omega)\left(B_1^2 + B_2^2 + B_3^2\right) - \frac{4}{3\mu}\left(A_1 B_1 + A_2 B_2 + A_3 B_3\right)$$
$$+\frac{4\lambda}{\mu}\left(A_1 A_2 + A_2 A_3 + A_1 A_3\right) + \frac{4\lambda}{3\mu}\left(B_1 B_2 + B_2 B_3 + B_1 B_3\right). \tag{A.1}$$

Varying this effective Lagrangian with respect to $A_n$, $B_n$, and $\mu$, we obtain a system of algebraic equations,

$$A_1^3 + (\omega-1)A_1 + \lambda(A_2 + A_3) - \frac{\mu}{3}B_1 + \frac{1}{5}A_1 B_1^2 = 0$$
$$A_2^3 + (\omega-1)A_2 + \lambda(A_1 + A_3) - \frac{\mu}{3}B_2 + \frac{1}{5}A_2 B_2^2 = 0$$
$$A_3^3 + (\omega-1)A_3 + \lambda(A_2 + A_1) - \frac{\mu}{3}B_3 + \frac{1}{5}A_3 B_3^2 = 0$$
$$\frac{3}{35}B_1^3 + \frac{1}{3}(\omega+1)B_1 + \frac{\lambda}{3}(B_2 + B_3) - \frac{\mu}{3}A_1 + \frac{1}{5}B_1 A_1^2 = 0 \tag{A.2}$$
$$\frac{3}{35}B_2^3 + \frac{1}{3}(\omega+1)B_2 + \frac{\lambda}{3}(B_1 + B_3) - \frac{\mu}{3}A_2 + \frac{1}{5}B_2 A_2^2 = 0$$
$$\frac{3}{35}B_3^3 + \frac{1}{3}(\omega+1)B_3 + \frac{\lambda}{3}(B_2 + B_1) - \frac{\mu}{3}A_3 + \frac{1}{5}B_3 A_3^2 = 0$$
$$\frac{1}{4}\left(A_1^4 + A_2^4 + A_3^4\right) + \frac{3}{140}\left(B_1^4 + B_2^4 + B_3^4\right) + \frac{1}{10}\left(A_1^2 B_1^2 + A_2^2 B_2^2 + A_3^2 B_3^2\right) + \frac{1}{2}(\omega-1)\left(A_1^2 + A_2^2 + A_3^2\right)$$
$$+\frac{1}{6}(1+\omega)\left(B_1^2 + B_2^2 + B_3^2\right) + \lambda(A_1 A_2 + A_2 A_3 + A_1 A_3) + \frac{\lambda}{3}(B_1 B_2 + B_2 B_3 + B_1 B_3) = 0.$$

## REFERENCES

1. R. Kashyap, *Fiber Bragg gratings* (Academic Press: San Diego, 1999).
2. H.G. Winful, J.H. Marburger, and E. Garmire, Appl. Phys. Lett. **35**, 379 (1979).
3. D.N. Christodoulides and R.I. Joseph, Phys. Rev. Lett. **62**, 1746 (1989).
4. A.B. Aceves and S. Wabnitz, Opt. Lett. **17**, 1566 (1989).
5. C.M de Sterke and J.E. Sipe, Prog. Opt. **33**, 203 (1994).

6. B.J. Eggleton, R.E. Slusher, C.M. de Sterke, P.A. Krug, and J.E. Sipe, Phys. Rev. Lett **76**, 1627 (1996); C.M. de Sterke, B.J. Eggleton, and P.A. Krug, J. Lightwave Technol. **15**, 1494 (1997).

7. B.A. Malomed and R.S. Tasgal, Phys. Rev. E **49**, 5787 (1994)

8. I.V. Barashenkov, D.E. Pelinovsky, E.V. Zemlyanaya, Phys. Rev. Lett. **80**, 5117 (1998).

9. A. De Rossi, C. Conti and S. Trillo, Phys. Rev. Lett. **81**, 85 (1998)

10. I.V. Barashenkov, E.V. Zemlyanaya, Comp. Phys. Comm. **126**, 22 (2000).

11. J. Schollmann, R. Scheibenzuber, A.S. Kovalev, A.P. Mayer, and A.A. Maradudin, Phys. Rev. E **59**, 4618 (1999).

12. W.C.K. Mak, P.L. Chu and B.A. Malomed, J. Opt. Soc. Am. B **15**, 1685 (1998).

13. A.V. Buryak and A.A. Akhmediev, J. Opt. Soc. Am. B **11**, 804 (1994).

14. A.B. Aceves and M. Santagiustina, Phys. Rev. E **56**, 1113 (1997).

15. R. Grimshaw, B.A. Malomed, and G.A. Gottwald, Phys. Rev. E **65**, 066606 (2002).

16. M. Johansson, e-print PS/0307057.

17. B.J. Eggleton, A.K. Ahuja, K.S. Feder, C. Headley, C. Kerbage, M.D. Mermelstein, and J.A. Rogers, P. Steinvurzel, P.S. Westbrook, and R.S. Windeler, IEEE J. Sel. Top. Quant. Electr. **7**, 409 (2001).

18. M. Åslund, L. Poladian, J. Canning J, and C. M. de Sterke, J. Lightwave Tech. **20**, 1585 (2002).

19. T.M. Monro and D.J. Richardson, Comptes Rendus Physique **4**, 175 (2003).

20. W.N. MacPherson, J.D.C. Jones, B.J. Mangan, J.C. Knight, and P.St.J. Russell, Opt. Commun. **223**, 375 (2003).

21. M.E. Potter and R.W. Ziolkowski, Opt. Exp. **10**, 691 (2002); M.C. Parker and S.D. Walker, IEEE J. Sel. Top. Quant. Elect. **8**, 1158 (2002).

22. V. Finazzi, T.M. Monro, and D.J. Richardson, J. Opt. Soc. Am. B **20**, 1427 (2003).

23. R.E. Slusher, S. Spalter, B.J. Eggleton, S. Pereira, and J.E. Sipe, Opt. Lett. **25**, 749 ( 2000).

24. A.V. Yulin, D.V. Skryabin, and W.J. Firth, Phys. Rev. E 66, 046603 (2002).

25. J. Yang, B.A. Malomed, and D.J. Kaup, Phys. Rev. Lett. **83**, 1958 (1999).

26. A.R. Champneys, B.A. Malomed, J. Yang and D.J. Kaup, Physica D **152-153**, 340 (2001); J. Yang, B.A. Malomed, D.J. Kaup and A.R. Champneys, Mathematics and Computers in Simulation **56**, 585 (2001).

27. P.G. Kevrekidis, B.A. Malomed, D.J. Frantzeskakis, and A.R. Bishop, Phys. Rev. E **67**, 036614 (2003)

28. D.J. Kaup. T. Lakoba, and B.A. Malomed, J. Opt. Soc. Am. B **14**, 1199 (1997).

29. F. Stenger, "*Numerical Methods Based on Sinc and Analytic Functions*", Springer-Verlag 1993, New York

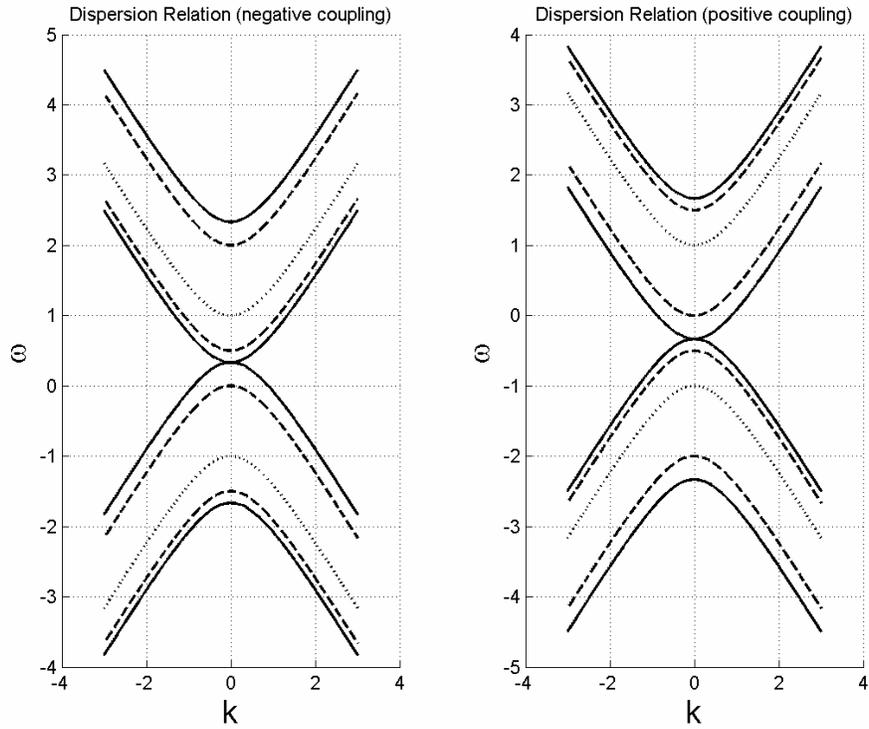

Fig. 1 Dispersion curves for the tri-core system. (a) Zero or negative values of the inter-core coupling constant, $\lambda = 0$ (dottted), $\lambda = -0.5$ (dashed), and $\lambda = -2/3$ (solid). (b) Zero or positive values of the coupling constant: $\lambda = 0$ (dotted), $\lambda = 0.5$ (dashed), and $\lambda = 2/3$ (solid).

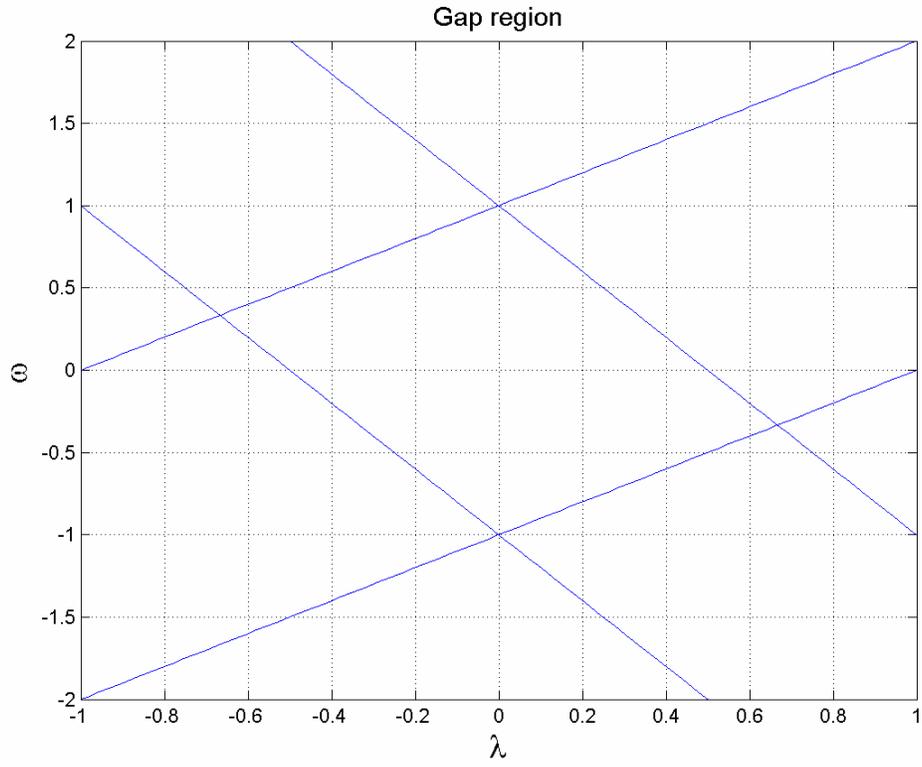

Fig. 2. The region in the parametric plane ($\lambda,\omega$) of the system of three linearly coupled Bragg gratings where the spectral gap exists. Two stripes are defined by Eqs. (3.3) and (3.4). The genuine gap is the rhombus formed by intersection of the stripes.

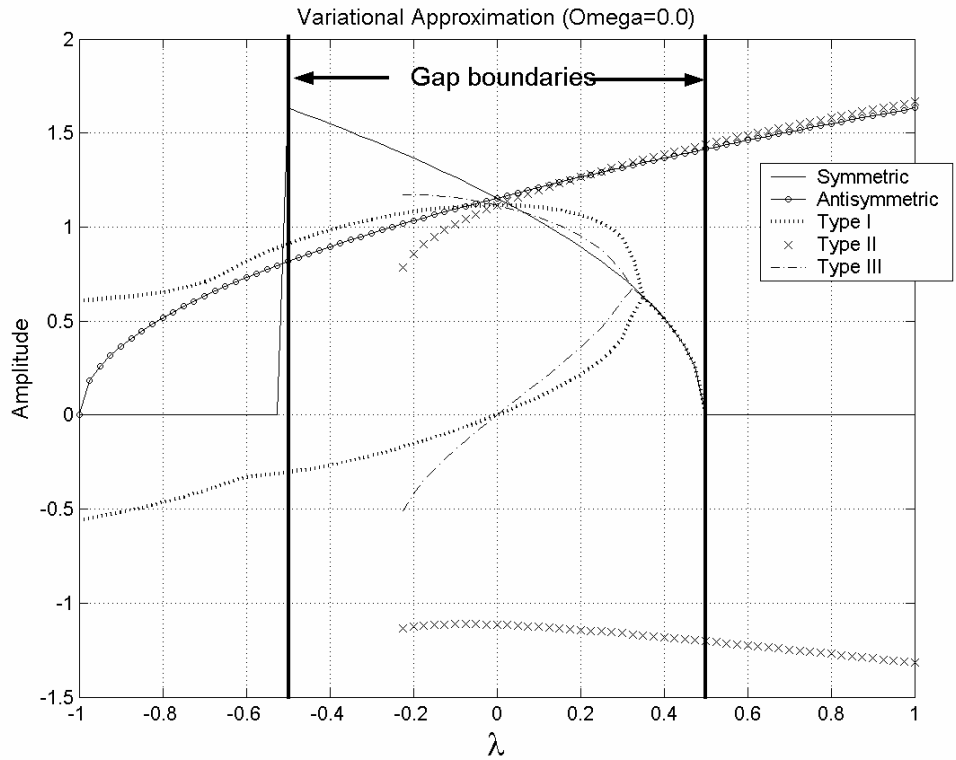

Fig. 3. Amplitudes of different types of solitons, found by means of the variational method, vs. the inter-core solitons $\lambda$ in the case of $\omega = 0$. Note that each symbol forms two different curves, corresponding to amplitudes of two different components of the soliton.

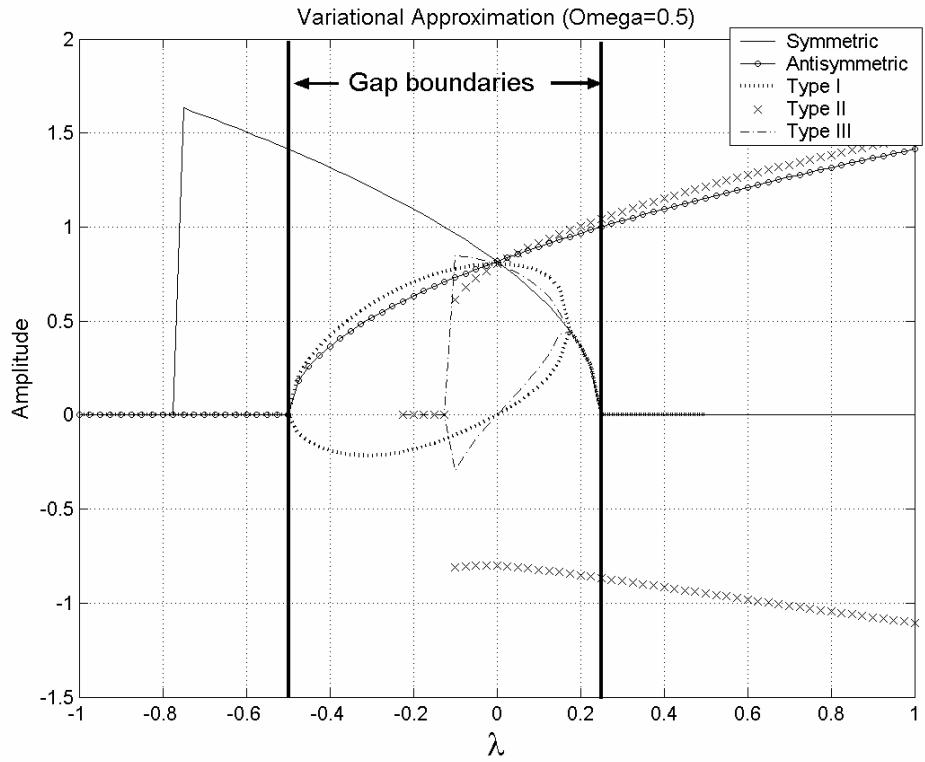

Fig. 4. The same as in Fig. 3, but for $\omega = 0.5$.

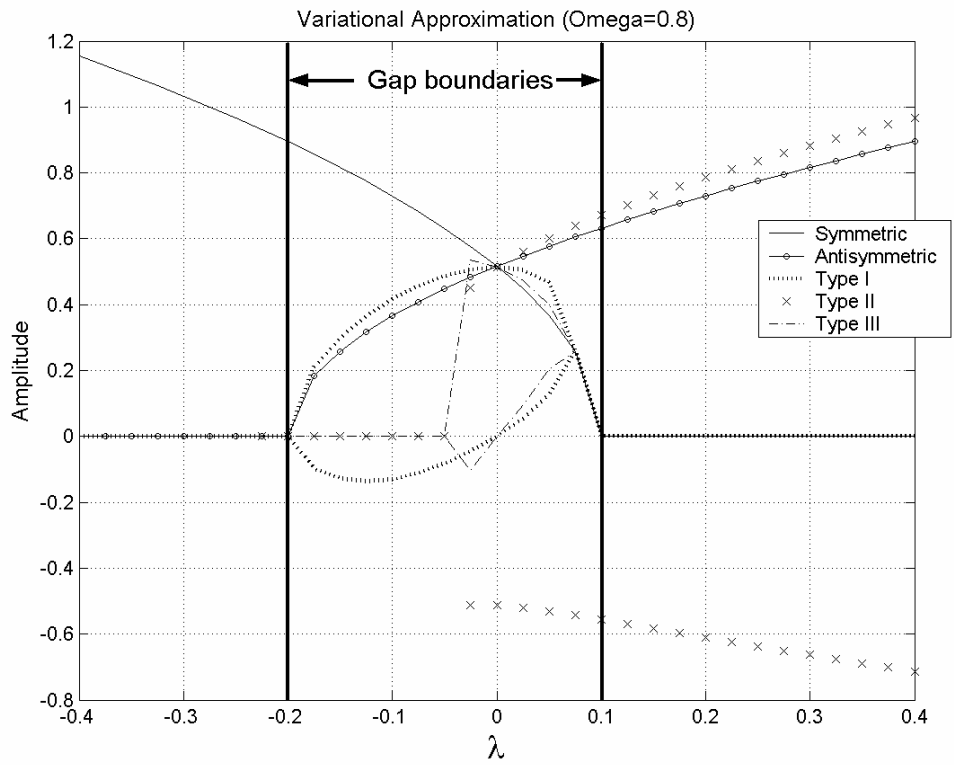

Fig. 5. The same as in Fig. 3, but for $\omega = 0.8$.

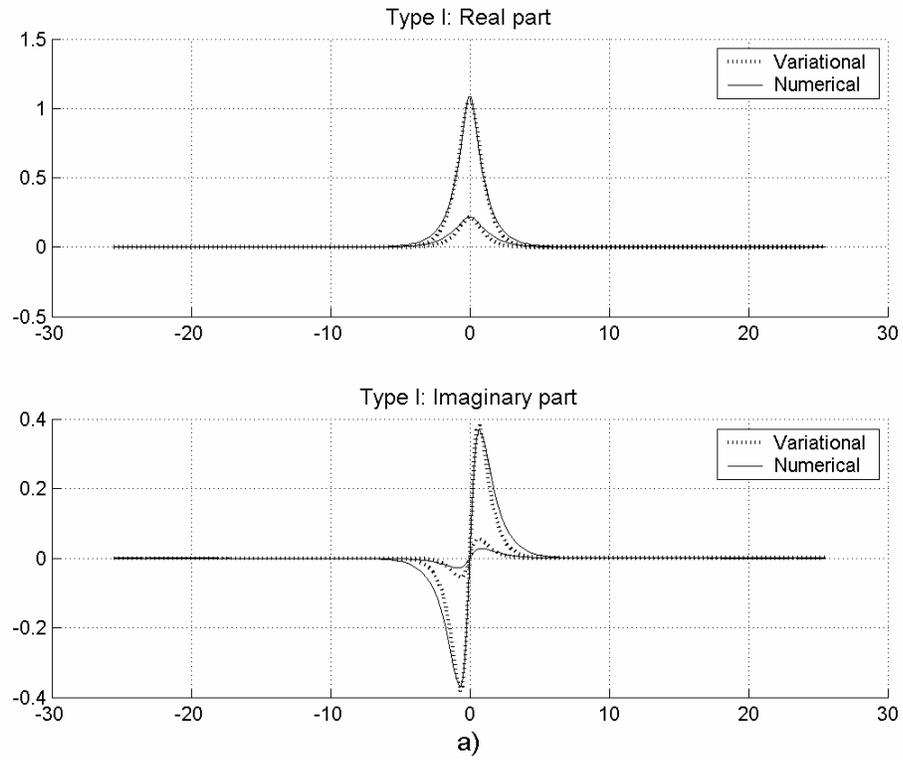

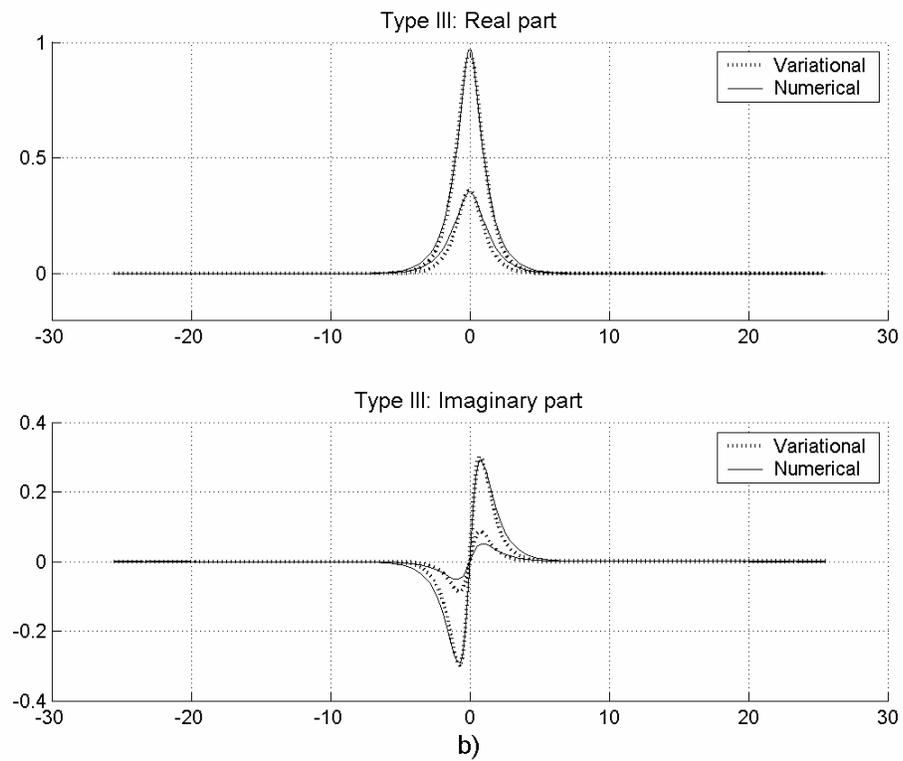

Fig. 6. Typical examples of asymmetric solitons, as found by means of the variational approximation (dashed) and in the numerical form (solid) for $\omega = 0$ and $\lambda = 0.2$: (a) type I; (b) type III.

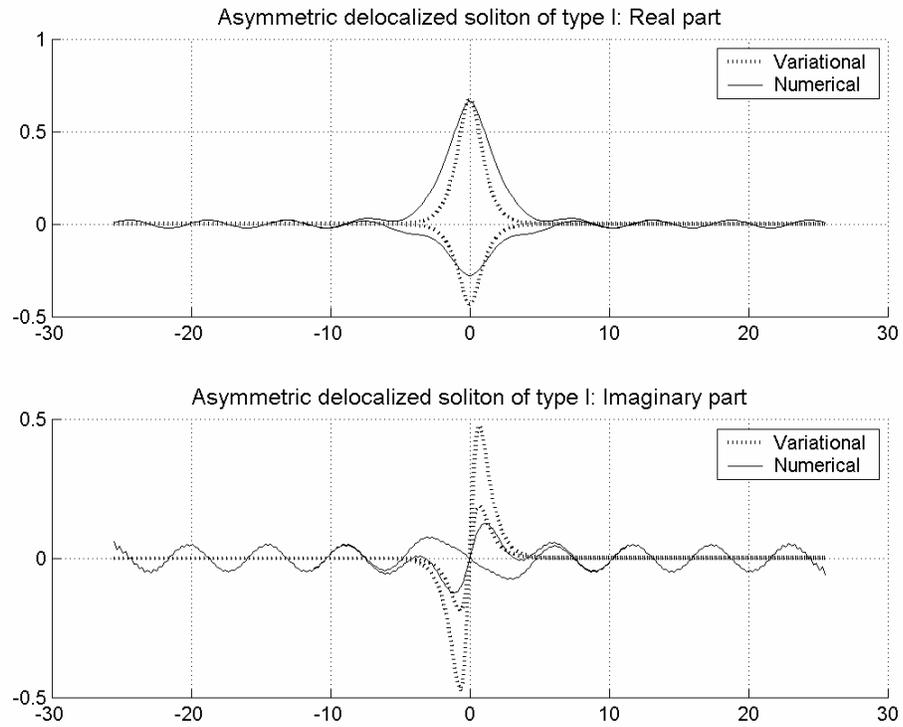

Fig. 7. A delocalized asymmetric soliton of type I, as found by means of the variational approximation (dashed) and by the numerical method (solid) for $\omega = 0$ and $\lambda = -0.75$.

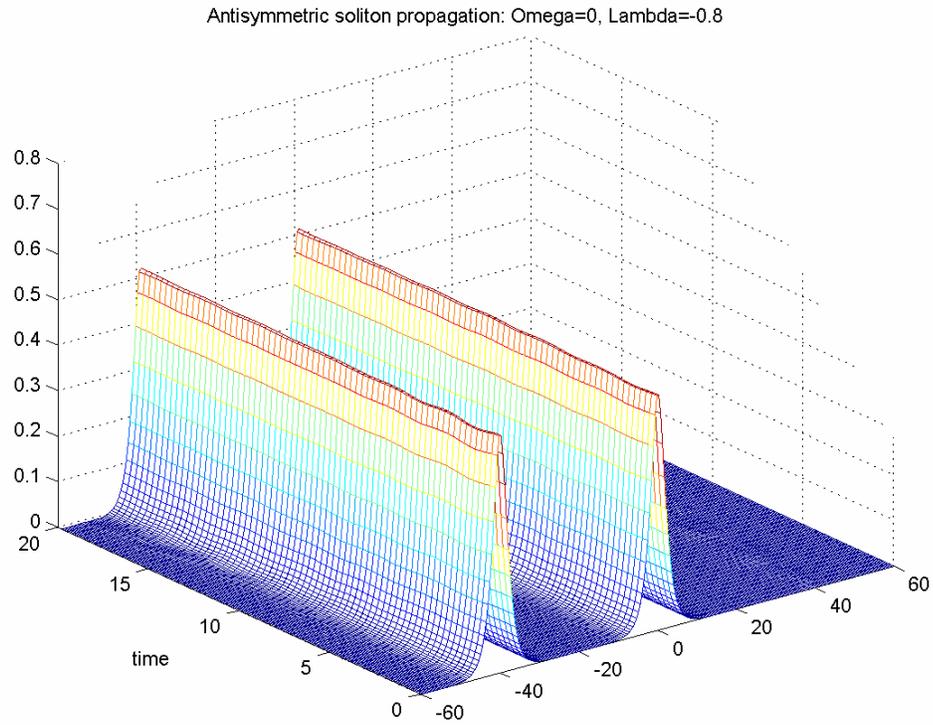

Fig. 8. A typical example of the stable antisymmetric soliton (with one empty core), found *outside* the spectral gap. Three components of the soliton are shown in one panel for the sake of compactness. Small intrinsic oscillations of the soliton observed in the figure are due to initial perturbations added to the soliton.

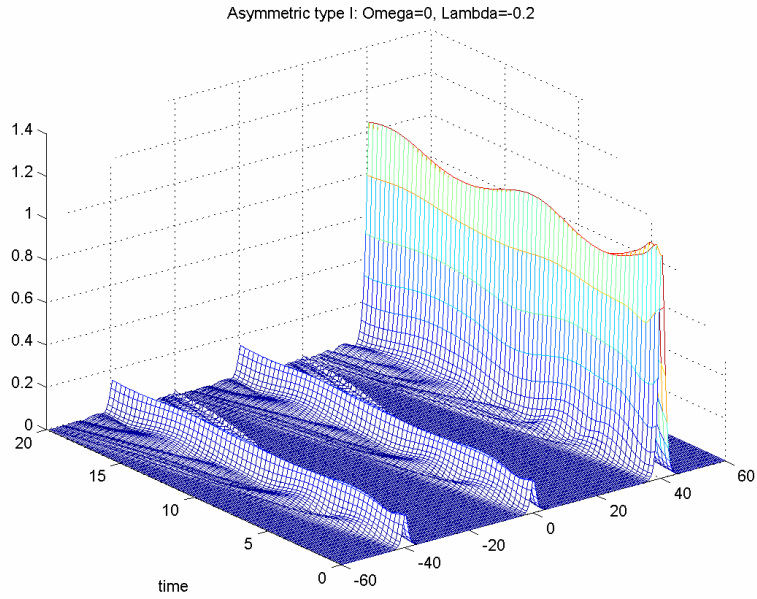

a)

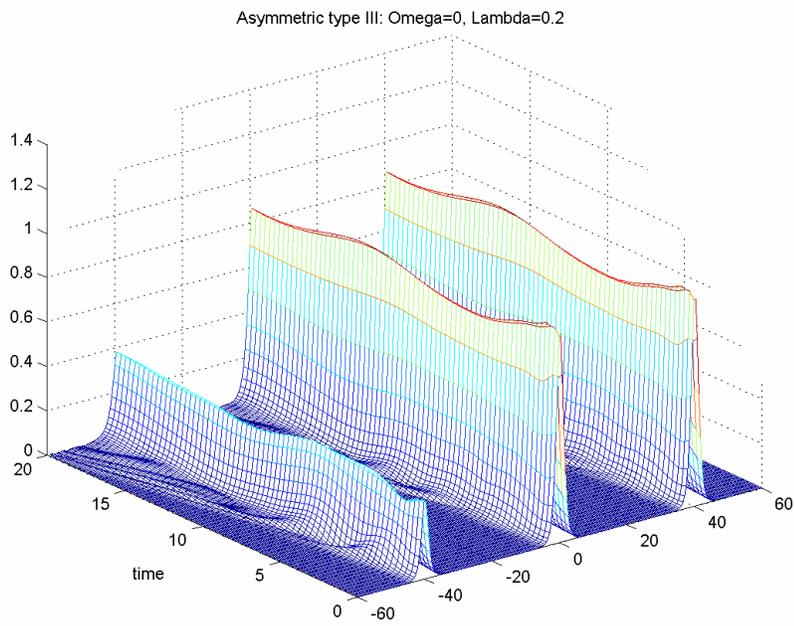

b)

Fig. 9. Evolution of stable asymmetric solitons: (a) type I; (b) type III. Three components of the solitons are shown in one panel for the compactness. Small intrinsic oscillations of the solitons observed in the figure are due to initial perturbations.

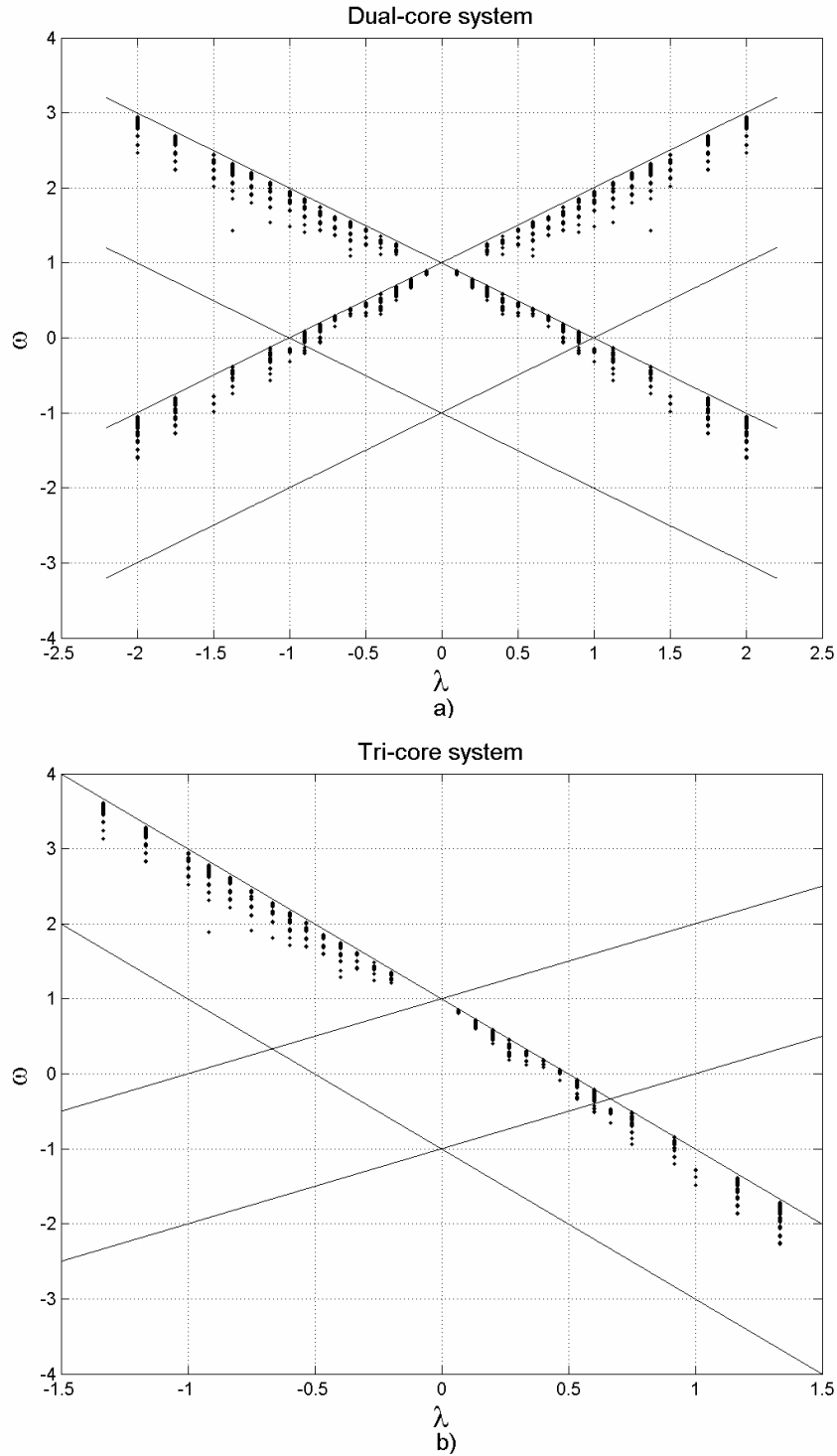

Fig. 10. The stability region of symmetric solitons: dots mark parameter sets ($\omega,\lambda$) for which it was checked that all the eigenvalues of small perturbations around the soliton are stable: (a) the dual-core system (this panel also includes the stability region for the antisymmetric solitons); (b) the tri-core system. The parallel lines with the negative and positive slope are existence borders for the symmetric and antisymmetric solitons, respectively.

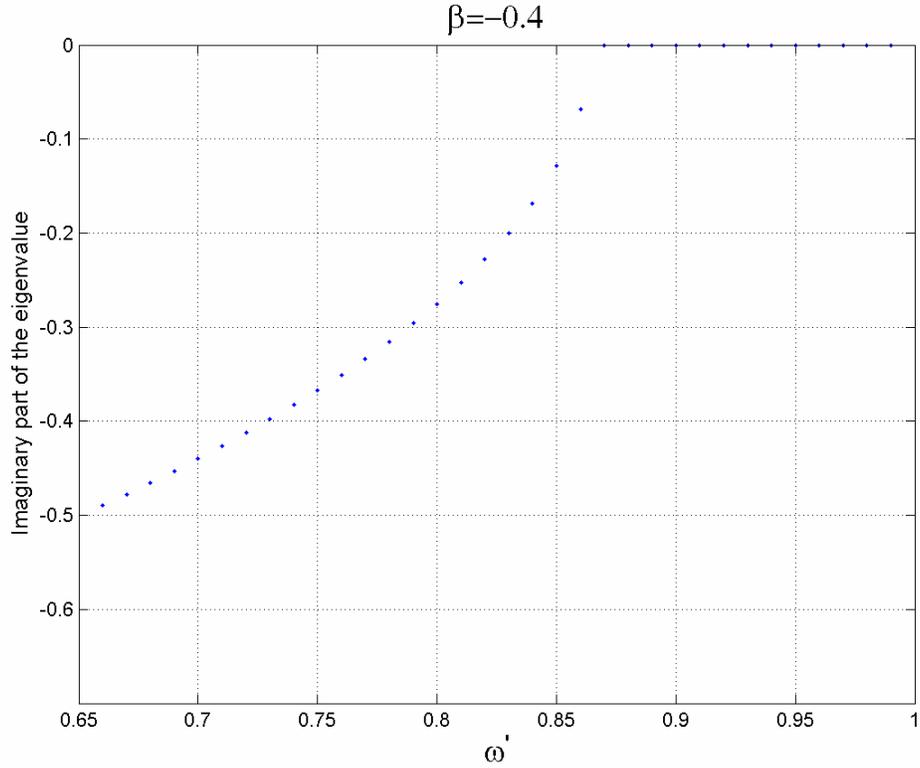

Fig. 11. A generic example of the appearance and evolution of an unstable eigenvalue in the spectrum of small perturbations around the symmetric soliton (for $\lambda = 0.2$) in the dual-core system. The horizontal axis is the frequency $\omega$ of the unperturbed soliton, redefined as per Eq. (6.10), and the absolute value of the imaginary part of the eigenvalue, shown on the vertical axis, is the growth rate of the oscillatory instability. The onset and subsequent development of the instability of symmetric solitons in the tri-core system are quite similar to those shown here and in Fig. 12 below.

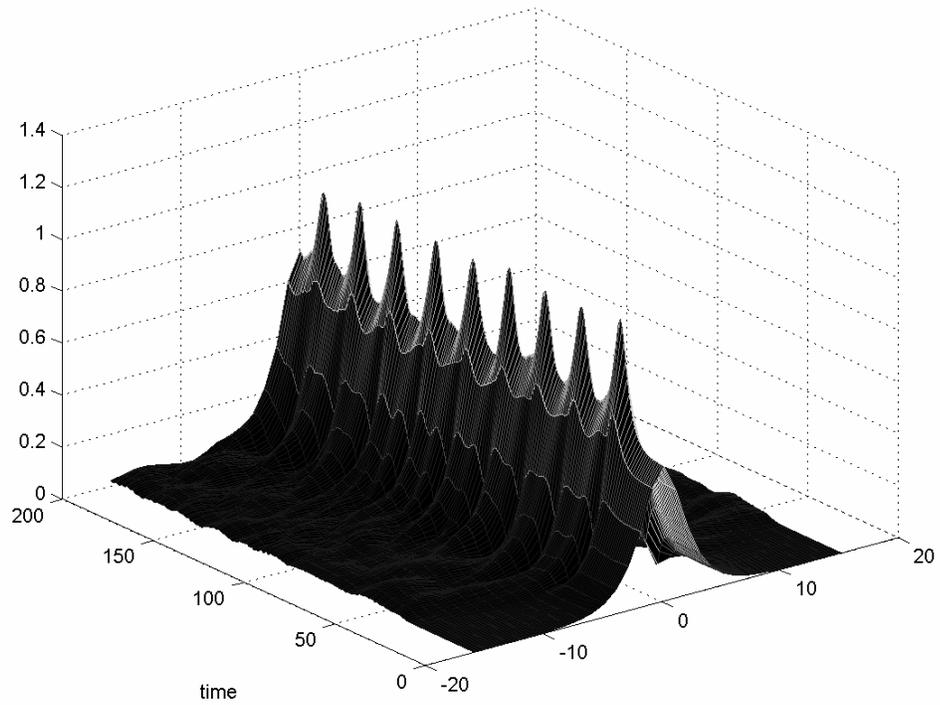

Fig. 12. Transformation of an unstable symmetric soliton into a breather by the oscillatory instability in the dual-core system, in the case of $\omega = 0.6$, $\lambda = 0.2$. After a transient period, the vibrations settle down to a quasi-steady state.